\documentclass[11pt]{article}
\usepackage[utf8]{inputenc}
\usepackage{booktabs}
\usepackage{hyperref}
\usepackage{url}
\usepackage{hyperref}
\usepackage{threeparttable}
\usepackage{graphicx}
\usepackage{float}

\usepackage{natbib}
\usepackage{graphicx}

\usepackage{color}
\usepackage[usenames,dvipsnames]{xcolor}
\usepackage{url}
\usepackage{hyperref}
\usepackage{color}
\definecolor{darkblue}{rgb}{0.0, 0.0, 0.5}
\definecolor{darkred}{rgb}{0.8, 0.0, 0.0}

\hypersetup{colorlinks=true,citecolor=darkblue,linkcolor=darkblue}

\begin{document}

\def\aj{AJ}%
\def\actaa{Acta Astron.}%
\def\araa{ARA\&A}%
\def\apj{ApJ}%
\def\apjl{ApJ}%
\def\apjs{ApJS}%
\def\ao{Appl.~Opt.}%
\def\apss{Ap\&SS}%
\def\aap{A\&A}%
\def\aapr{A\&A~Rev.}%
\def\aaps{A\&AS}%
\def\azh{AZh}%
\def\baas{BAAS}%
\def\bac{Bull. astr. Inst. Czechosl.}%
\def\caa{Chinese Astron. Astrophys.}%
\def\cjaa{Chinese J. Astron. Astrophys.}%
\def\icarus{Icarus}%
\def\jcap{J. Cosmology Astropart. Phys.}%
\def\jrasc{JRASC}%
\def\mnras{MNRAS}%
\def\memras{MmRAS}%
\def\na{New A}%
\def\nar{New A Rev.}%
\def\pasa{PASA}%
\def\pra{Phys.~Rev.~A}%
\def\prb{Phys.~Rev.~B}%
\def\prc{Phys.~Rev.~C}%
\def\prd{Phys.~Rev.~D}%
\def\pre{Phys.~Rev.~E}%
\def\prl{Phys.~Rev.~Lett.}%
\def\pasp{PASP}%
\def\pasj{PASJ}%
\def\qjras{QJRAS}%
\def\rmxaa{Rev. Mexicana Astron. Astrofis.}%
\def\skytel{S\&T}%
\def\solphys{Sol.~Phys.}%
\def\sovast{Soviet~Ast.}%
\def\ssr{Space~Sci.~Rev.}%
\def\zap{ZAp}%
\def\nat{Nature}%
\def\iaucirc{IAU~Circ.}%
\def\aplett{Astrophys.~Lett.}%
\def\apspr{Astrophys.~Space~Phys.~Res.}%
\def\bain{Bull.~Astron.~Inst.~Netherlands}%
\def\fcp{Fund.~Cosmic~Phys.}%
\def\gca{Geochim.~Cosmochim.~Acta}%
\def\grl{Geophys.~Res.~Lett.}%
\def\jcp{J.~Chem.~Phys.}%
\def\jgr{J.~Geophys.~Res.}%
\def\jqsrt{J.~Quant.~Spec.~Radiat.~Transf.}%
\def\memsai{Mem.~Soc.~Astron.~Italiana}%
\def\nphysa{Nucl.~Phys.~A}%
\def\physrep{Phys.~Rep.}%
\def\physscr{Phys.~Scr}%
\def\planss{Planet.~Space~Sci.}%
\def\procspie{Proc.~SPIE}%
\let\astap=\aap
\let\apjlett=\apjl
\let\apjsupp=\apjs
\let\applopt=\ao
\def\uol{University of Louisville}
\def\uct{University of Cape Town}
\def\rutgers{Rutgers University, The State University of New Jersey}
\def\stsci{Space Telescope Science Institute}
\def\stsci{Space Telescope Science Institute}
\def\icrar{ICRAR}
\def\uwc{UWC/IDIA}
\def\unc{University of North Carolina, Chapel Hill}
\def\wisconsin{University of Wisconsin-Madison}

\def\adler{Adler Planetarium, Chicago, IL, USA}
\def\aims{African Institute for Mathematical Sciences, 6 Melrose Road, Muizenberg 7945, South Africa}
\def\arizona{University of Arizona, Tucson, AZ, USA}
\def\berkeley{Physics Division, Lawrence Berkeley National Laboratory, 1 Cyclotron Road, Berkeley, CA, 94720, USA}
\def\caltech{California Institute of Technology, Pasadena, CA, USA}
\def\caltechgps{Division of Geological and Planetary Sciences, California Institute of Technology, Pasadena, CA 91125, USA}
\def\centrallancashire{University of Central Lancashire, Preston PR1 2HE, UK}
\def\cfa{Harvard-Smithsonian Center for Astrophysics, Harvard University, Cambridge, MA, USA}
\def\chicago{Department of Astronomy and Astrophysics, University of Chicago, 5640 South Ellis Avenue, Chicago, IL 60637, USA}
\def\cmu{Carnegie Mellon University, Pittsburgh, PA, USA}
\def\columbia{Columbia University, New York, NY, USA}
\def\cook{Cook Astronomical Consulting, USA}
\def\cuny{The City University of New York, New York, NY, USA}
\def\ctio{Cerro Tololo Inter-American Observatory, Casilla 603, La Serena, Chile}
\def\dearborn{University of MichiganâDearborn, 4901 Evergreen Road, Dearborn, MI 48128, USA}
\def\delaware{University of Delaware, Department of Physics and Astronomy, 104 The Green, Newark, DE 19716, USA}
\def\drexel{Drexel University, Philadelphia, PA, USA}
\def\ennu{Department of Physics \& Astronomy/CIERA, Northwestern University, 2145 Sheridan Road, Evanston, IL, 60208, USA}
\def\fermilab{Fermilab, PO Box 500, Batavia, IL, 60510, USA}
\def\floridagulf{Florida Gulf Coast University, Fort Meyers, FL, USA}
\def\goddard{NASA Goddard Space Flight Center, 8800 Greenbelt Road, Greenbelt, MD 20771, USA}
\def\harvard{Department of Physics \& Department of Astronomy, 17 Oxford Street, Harvard University, Cambridge, MA, 02138, USA}
\def\ifa{Institute for Astronomy, University of Hawaii at Manoa, 2680 Woodlawn Drive, Honolulu, HI 96822, USA}
\def\ipac{IPAC, 770 South Wilson Ave., Pasadena, CA 91125, USA}
\def\irvine{University of California, Irvine, CA, USA}
\def\jpl{Jet Propulsion Laboratory, California Institute of Technology, 4800 Oak Grove Drive, Pasadena, CA 91109, USA}
\def\kicp{Kavli Institute for Cosmological Physics, University of Chicago, Chicago, IL 60637, USA}
\def\lcogt{LCOGT, University of California, Santa Barbara, CA, USA}
\def\lpc{Laboratoire de Physique de Clermont, N2P3/CNRS, 63178 AubiÃšre Cedex, France}
\def\lpnhe{LPNHE, Barre 12-22, 1er s\'{e}tage, 4 Place Jussieu, 75252 Paris Cedex 05, France}
\def\lsst{LSST, 933 N. Cherry Ave., Tucson, AZ 85721, USA}
\def\mssl{Mullard Space Science Laboratory (MSSL), University College London (UCL), Surrey RH5 6NT, UK}
\def\msu{Department of Physics and Astronomy, Michigan State University, 5678 Wilson Road, Lansing, MI 48824, USA}
\def\nau{Dept. of Physics \& Astronomy, Northern Arizona University, NAU Box 6010, Flagstaff, AZ, 86011, USA}
\def\nso{National Solar Observatory, 3004 Telescope Loop, Sunspot, NM 88349, USA}
\def\noao{NOAO, 950 N. Cherry Ave., Tucson, AZ 85719}
\def\nyu{Center for Cosmology and Particle Physics, Department of Physics, New York University, 726 Broadway, 9th Floor, New York, NY 10003, USA}
\def\okc{The Oskar Klein Centre for Cosmoparticle Physics, Stockholm University, Stockholm, Sweden}
\def\nyuc{Center for Cosmology and Particle Physics, Department of Physics, New York University, 726 Broadway, 9th Floor, New York,  NY 10003, USA}
\def\osu{The Ohio State University, Columbus, OH, USA}
\def\oswego{State University of New York at Oswego, 7060 New York 104, Oswego, NY 13126, USA}
\def\oxford{Department of Physics, University of Oxford, Keble Road, Oxford, UK}
\def\penn{Department of Astronomy and Astrophysics, University of Pennsylvania, Philadelphia, PA, USA}
\def\pennstate{Pennsylvania State University, 514A Davey Lab
University Park, PA 16802, USA}
\def\pitt{Pittsburgh Particle Physics, Astrophysics, and Cosmology Center (PITT PACC), Physics and Astronomy Department, University of Pittsburgh, Pittsburgh, PA 15260, USA}
\def\princeton{Department of Astrophysical Sciences, Princeton University, Princeton, NJ 08544, USA}
\def\rice{Department of Physics and Astronomy, Rice University, Houston TX 77005-1892, USA}
\def\rutgers{Department of Physics and Astronomy, Rutgers the State University of New Jersey, 136 Frelinghuysen Road, Piscataway, NJ 08854 USA}
\def\uc{Department of Astronomy and Astrophysics, University of California, Santa Cruz, CA 95064, USA}
\def\scsu{Department of Physics, Southern Connecticut State University, 501 Crescent Street, New Haven, CT 06515, USA}
\def\seti{SETI Institute, 189 N. Bernardo Ave., Mountain View, CA, 94043, USA}
\def\ska{SKA South Africa, 3rd Floor, The Park, Park Road, Pinelands 7405, South Africa}
\def\sofia{SOFIA Science Center, NASA Ames Research Center, MS211-1, Moffett Field, CA 94035, USA}
\def\slac{SLAC National Accelerator Laboratory, 2575 Sand Hill Road, MS29, Menlo Park, CA 94025, USA}
\def\somewhere{Some Institute, Somewhere, \ldots}
\def\soton{School of Physics and Astronomy, University of Southampton, Southampton, SO17 1BJ, UK}
\def\stanford{Physics Department, Stanford University, Stanford, CA, 94305, USA}
\def\stsci{Space Telescope Science Institute, Baltimore, MD, USA}
\def\texastech{Department of Physics, Texas Tech University, Box 41051 Lubbock, TX 79409-1051, USA}
\def\toronto{Dunlap Institute \& Department of Astronomy and Astrophysics, University of Toronto, 50 St George Street, Toronto, ON M5S 3H4, Canada}
\def\ucd{University of California, Davis, CA, USA}
\def\ucl{Department of Physics and Astronomy, University College London, Gower Street, London WC1E 6BT, UK}
\def\unab{Universidad Andr\'{e}s Bello, 13 Nte. 798, Vi\~{n}a del Mar, RegiÃ³n de ValparaÃ­so, Chile}
\def\unt{University of North Texas, 1155 Union Cir, Denton, TX 76203, USA}
\def\usno{US Naval Observatory, 10391 West Naval Observatory Road, Flagstaff, AZ 86001, USA}
\def\utaustin{University of Texas at Austin,  Austin, TX, 78712, USA}
\def\uw{University of Washington, Department of Astronomy, University of Washington, 3910 15th Avenue NE, Seattle, WA, 98195, USA}
\def\uwe{The eScience Institute, University of Washington, Seattle, WA, 98195, USA}
\def\westernwash{Western Washington University, 516 High Street, Bellingham, WA 98225, USA}
\def\vanderbilt{Vanderbilt University, 2201 West End Ave, Nashville, TN 37235, USA}
\def\yale{Department of Astronomy, Yale University, P.O. Box 208101, New Haven, CT 06520-8101, USA}

\def\hi{H{\sc I}}
\title{Large Synoptic Survey Telescope White Paper;\\ 
The Case for Matching U-band on Deep Drilling Fields 
}

\author{B.W. Holwerda (University of Louisville), \\
A. Baker (Rutgers University), \\
S. Blyth (University of Cape Town), \\
S. Kannappan (University of North Carolina), \\
D. Obreschkow (ICRAR), \\
S. Ravindranath (STSCI), \\
E. Elson (University of Western Cape), \\
M. Vaccari (University of Western Cape), \\
S. Crawford (STSCI), \\
M. Bershady (University of Wisconsin-Madison), \\
N. Hathi (STSCI) \\
N. Maddox (Ludwig Maximilians University Munich) \\
R. Taylor (University of Cape Town) \\
M. Jarvis (University of Oxford)\\
J. Bridge (University of Louisville)
}

\date{November 2018}

\maketitle

\begin{abstract}
U-band observations with the LSST have yet to be fully optimized in cadence. The straw man survey design is a simple coverage of the medium-deep-fast survey. Here we argue that deep coverage of the four deep drilling fields (XMM-LSS, ECDFS, ELAIS-S1 and COSMOS) has a much higher scientific return, given that these are also the target of the Southern Hemisphere's Square Kilometer Array Pathfinder, the MeerKAT specifically, deep radio observations.
\end{abstract}

\section{White Paper Information}
The point of contact for this white paper is Benne W. Holwerda \url{benne.holwerda@louisville.edu} (\uol).

\begin{enumerate} 
\item {\bf Science Category:} Galaxies
\item {\bf Survey Type Category:} Deep Drilling field
\item {\bf Observing Strategy Category:} a specific pointing or set of pointings that is (relatively) agnostic of the detailed observing strategy or cadence. See \cite{Brandt18} for an example on the cadence requirements for extra-galactic science.
\end{enumerate}  

\clearpage

\section{Scientific Motivation}


\subsection{Introduction}
\label{s:intro}

The Large Synoptic Survey Telescope (LSST) collaboration issues a call for white papers regarding the refinement of the cadence strategy, with an eye to improve possible synergy science with other world-class instruments. In this paper, we focus on the deep drilling fields, specifically the need to include {\em u-band} LSST observations of comparable depth as the stack of the remaining LSST filters.

U-band observations reveal the recent massive star-formation in galaxies in the deep drilling fields. In combination with continuum, and especially \hi\ observations by the new Square Kilometer Array \citep[SKA,][]{Blyth15} coming online concurrently or just before LSST operations. The survey strategy for the southern SKA Pathfinders, MeerKAT and ASKAP, has mostly been settled into several tiers of radio continuum and \hi\ surveys, coinciding in many places with the LSST deep drilling fields.
We therefore identify an opportunity to maximize scientific return by adding u-band observations to the LSST deep fields to reap the benefits.


\begin{figure}[h]
  \begin{center}
    \begin{minipage}[lt]{\linewidth}
	\begin{center}
 	\includegraphics[width=\textwidth]{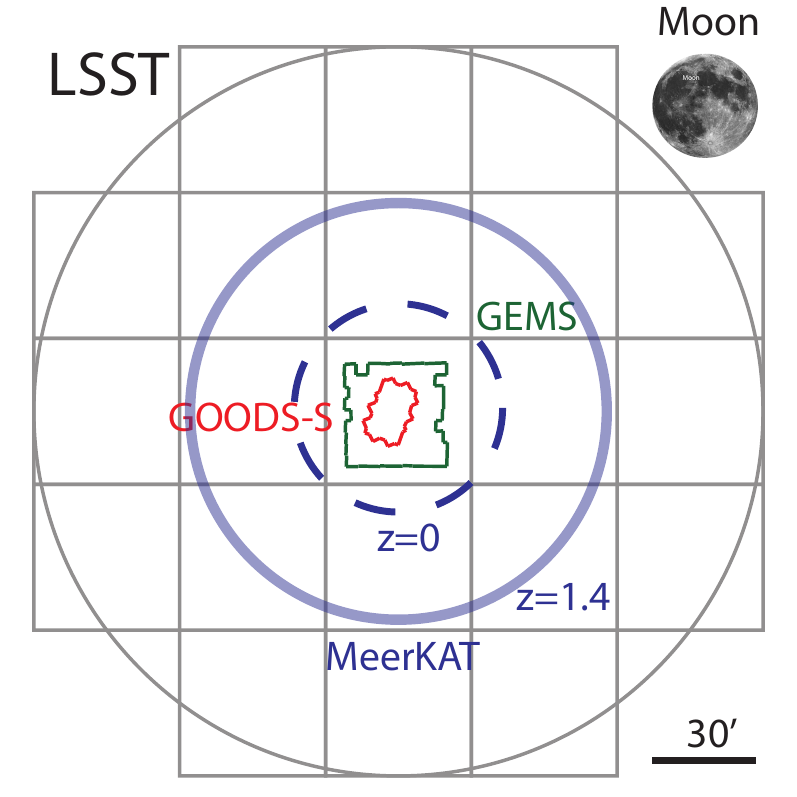}
		\caption{\label{f:fov1}The field-of-view of the LADUMA survey at z=0 and z=1.4 (blue dashed and solid line respectively), the GEMS (green) and GOODS/CANDELS (red) coverage with HST, and an LSST FOV (gray).  } 
         \end{center}
     \end{minipage}\hfill
   \end{center}
\end{figure}

\subsection{SKA Pathfinder Science}
\label{s:skascience}

The SKA Pathfinders in the Southern Hemisphere, the South African MeerKAT \citep[Karoo Array Telescope][]{meerkat1,meerkat2,MeerKAT} and the Australian ASKAP \citep[Australian SKA Pathfinder][]{askap1,askap2,askap3,askap4,ASKAP} have iterated to an effective tiered \hi\ and radio continuum survey strategy. For the galaxy evolution science cases for the LSST deep drilling fields the deepest two tiers are the most relevant: the deepest LADUMA \hi\ survey and the MIGHTEE medium deep \hi\ and continuum survey  with MeerKAT. 

\begin{figure}[h]
  \begin{center}
    \begin{minipage}[lt]{\linewidth}
	\begin{center}
    		\includegraphics[width=\textwidth]{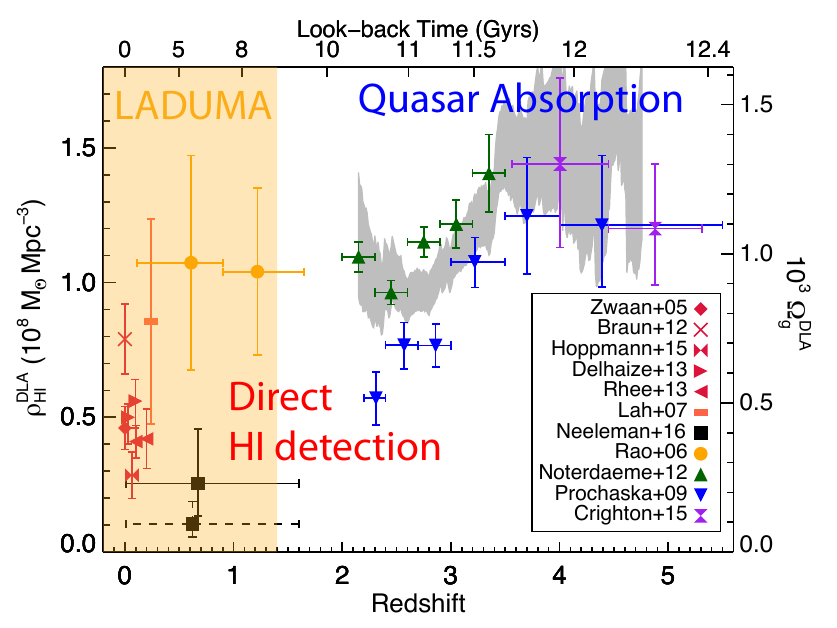}
		\caption{\label{f:OmHI}($\Omega_{HI}$) -- The density of H\,{\sc{i}} gas in the Universe as a function of redshift (or look-back time, top x-axis) from \protect\cite{Neeleman16}. High-redshift measurements (quasar absorption) are paradoxically more accurate than lower redshift direct detection ones.  Lower-redshift results (mostly from direct observation of hydrogen's 21cm emission) remain uncertain, e.g., the one order of magnitude mismatch between \protect\cite{Rao06} and \protect\cite{Neeleman16}. LADUMA will resolve the evolution of cosmic H\,{\sc{i}} supply, when star-formation density drops and cosmic gas supply is critical to understand (z=0-1.5). }
         \end{center}
     \end{minipage}\hfill
   \end{center}
\end{figure}

\subsubsection{The LADUMA Survey}
\label{s:laduma}

To trace the fueling of galaxies over the latter half of the Age of the Universe, LADUMA was proposed in 2009 to be the deepest ($\sim$5000 hours) integration with a radio telescope to date. The LADUMA survey\footnote{PIs B.W. Holwerda, S-L. Blyth, and A. Baker \& 71 co-Is, see: \url{ http://www.laduma.uct.ac.za/} ``Laduma" means ``goal!" in Zulu, one of the national languages of South Africa, appropriate during the 2010 soccer world cup.} has been designated top-two priority of the eight guaranteed-time large projects for the MeerKAT radio telescope.
Starting full science operations in 2019, it will observe the Chandra Deep Field South to detect \hi\ emission from a Milky Way analog out to $z\sim1.4$ \citep[7 Gyr][]{Holwerda10vuvu, Holwerda11aas,Blyth15a,Baker18}. Given the current foreseen roll-out of SKA capabilities, the depth and redshift range of the LADUMA \hi\ data-cube is going to be unsurpassed until SKA-2 comes online ($>$2030). The LADUMA survey represents the deepest tier in a series of new surveys aimed at characterizing  gas in galaxies. Direct detection of $M_{HI}^*$ are expected for the entire volume of the survey: initially out to $z\sim0.6$ and later to z=1.4 once the second set of receivers is installed.
The main science cases for LADUMA are (i) the evolution of the \hi\ volume density in the Universe over cosmic time-scales, (ii) the evolution of galaxy kinematics and disk assembly (
i.e., the evolution of the Baryonic Tully-Fisher relation), (iii) the evolution of neutral gas reserves of galaxy populations, and (iv) the evolution of the neutral gas mass function (HIMF).

\begin{figure}[h]
  \begin{center}
    \begin{minipage}[lt]{\linewidth}
	\begin{center}
 	\includegraphics[width=\textwidth]{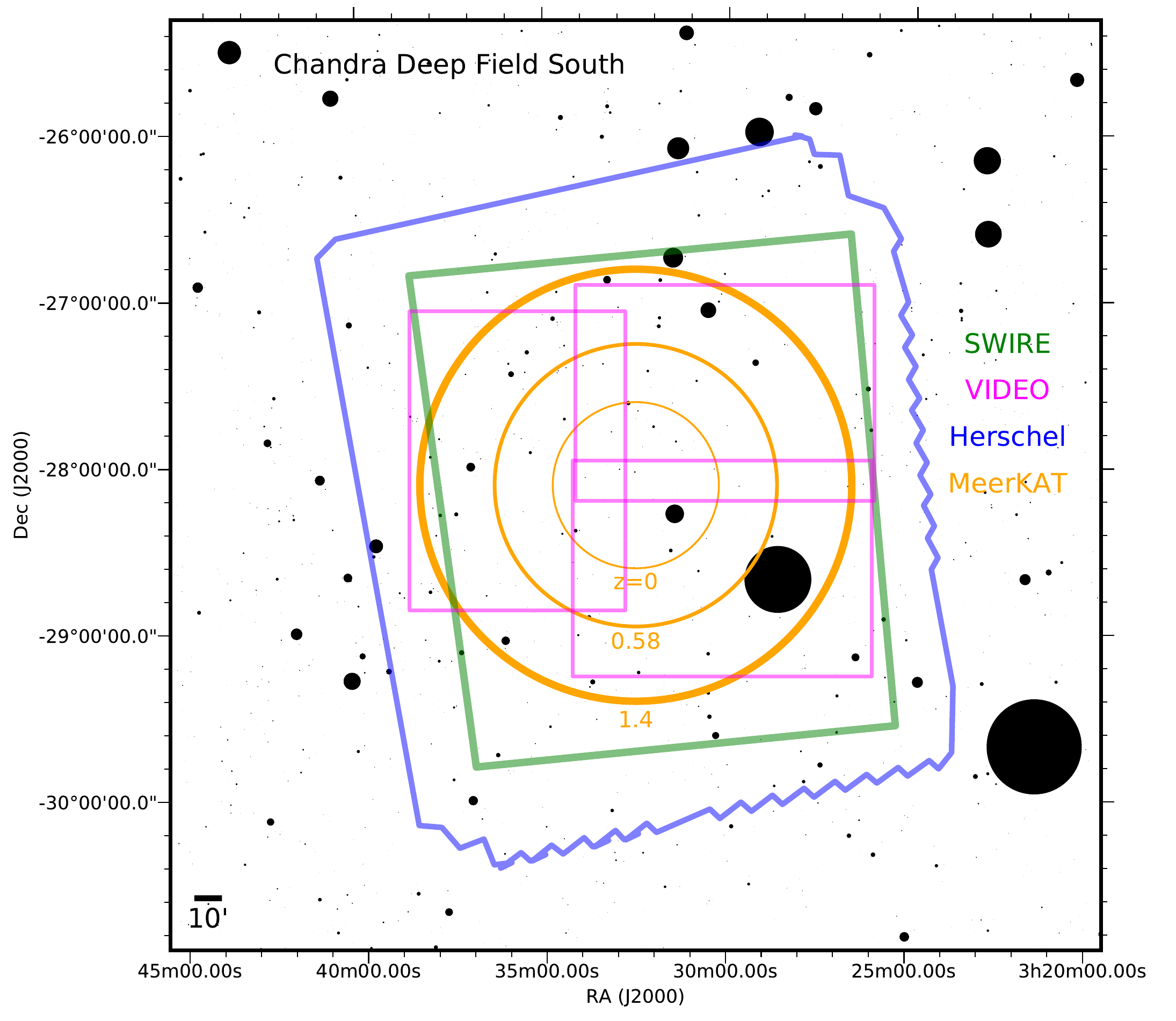}
		\caption{\label{f:fov2} An illustration of the wide range of multiwavelength coverage already available on the LADUMA target field (SDF-S). SWIRE (Spitzer), VIDEO (IR), Herschel, and the FOV of the MeerKAT radio array at the 21cm line emission of hydrogen at different redshifts. Points denote continuum sources identified with the VLA.} 
         \end{center}
     \end{minipage}\hfill
   \end{center}
\end{figure}

\subsubsection{MIGHTEE}
\label{s:mightee}

The MIGHTEE large survey project on the MeerKAT telescope \citep{mightee} will survey four of the most well-studied extragalactic deep fields, totaling 20 square degrees to $\mu$Jy radio continuum sensitivity at Giga-Hertz frequencies, as well as an ultra-deep image of a single $\sim1$ square degree MeerKAT pointing, commensally with LADUMA. The observations will provide radio continuum, spectral line (\hi), and polarisation information. The combination of these MIGHTEE data, together with existing and near-future multi-wavelength data (e.g. LSST observations) will allow a range of AGN and galaxy evolution science. 

Specifically, MIGHTEE is designed to significantly enhance our understanding of, 
(i) the evolution of AGN and star-formation activity over cosmic time, as a function of stellar mass and environment, free of dust obscuration; 
(ii) the evolution of neutral hydrogen (\hi) in the Universe to $z<0.4$ benefitting from a larger volume than LADUMA and how this neutral gas eventually turns into stars after moving through the molecular phase, and how efficiently this can fuel AGN activity; 
(iii) the properties of cosmic magnetic fields and how they evolve in clusters, filaments and galaxies. MIGHTEE will reach similar depth to the planned SKA all-sky survey, and thus will provide a pilot to the cosmology experiments that will be carried out by the SKA over a much larger survey volume.

MIGHTEE targets the XMM-LSS, ECDFS, ELAIS-S1 and COSMOS, which already coincide with the LSST deep drilling fields.

\subsection{Science Cases}
\label{s:science}

The LSST u-band predominantly traces massive star-formation, both in the nearby Universe as well as out to cosmic noon, the height of star-formation at $z\sim1$ \citep{Madau98,Madau14}, where it samples the near ultraviolet. Deep u-band photometry, especially in conjunction with \hi\ information allows one to probe the evolution in several revealing galaxy scaling relations.  

\begin{figure}[h]
  \begin{center}
    \begin{minipage}[lt]{\linewidth}
	\begin{center}
 	\includegraphics[width=\textwidth]{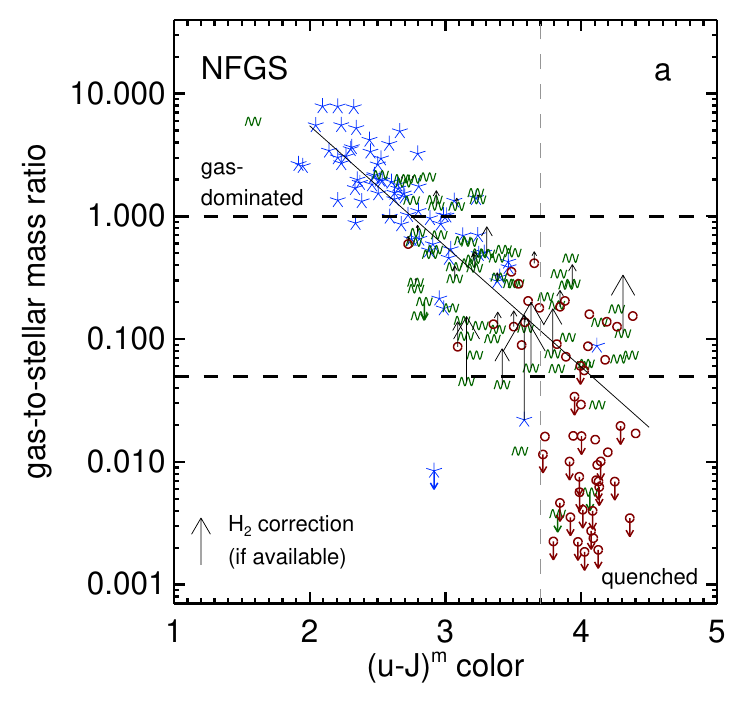}
\caption{\label{f:gasfrac}Gas richness ($M_{HI}/M^*$) in relation to U--J color from \cite{Kannappan13}, with symbols indicating morphology class: disks (blue stars), disk and bulge (green wave), and spheroidal (red circle). The solid line is a fit minimizing residuals in $M_{HI}/M^*$ for all galaxies bluer than (u--J) = 3.7, which yields $log(M_{HI}/M^*) = 2.70 + 0.98(u-J)$ with 0.34 dex scatter. }
         \end{center}
     \end{minipage}\hfill
   \end{center}
\end{figure}

\subsubsection{UV-IR vs. $M_{HI}/M^*$} 
\label{s:uvir}

The deep u-band observations with LSST combined with SKA Pathfinder \hi\ will 
(1) calibrate the redshifted photometric gas fraction relation (UV-IR vs. $M_{HI}/M^*$, currently calibrated only at z=0) and 
(2) use this relation with stellar population synthesis to understand galaxy star formation and gas accretion histories 
\citep[see][]{Kannappan04, Kannappan13, Eckert15}. The u-band and optical morphology can directly be compared to the SDSS and GALEX data for which this relation was determined. The UV-IR vs. $M_{HI}/M^*$ relation (Figure \ref{f:gasfrac}) quantifies how and how efficient galaxies convert gas into stars. With u-band and \hi\ observations in hand, the changes in these relations can directly be observed. 

\subsubsection{Star-formation and the gas reserve of galaxies} 
\label{s:sfrhi}

LSST U-band imaging is an excellent way to estimate the actual star-formation rate which contributes to the study of the gas content and fractions as function of SF activity. The u-band imaging is the best estimator to use here in combination with \hi\ fuel estimates. The \hi\ traces the diffuse ISM, predominant in the outer regions of galaxies while u-band traces the un-attenuated massive star-formation, a fraction of which happens in the outer disks. Galaxy-averaged comparisons can be done with LSST u-band, as well as quantify how it is distributed within a galaxy, a key issue. 

\subsubsection{Growth of galaxy disks} 
\label{s:diskgrowth}

The galaxy disk mass-size relation grows over cosmic time \citep{Shen03, Baldry12,van-der-Wel14,Shibuya15}. Disk scale-length depends on waveband (u-band vs optical and near-infrared) and it is expected galaxies grow at different rates in different wavebands, depending on gas acquisition \citep{Fall80,Mo98}. 

The triple threat of LSST morphologies can trace this growth of galaxy disks with gas reservoir: optical band ({\em gri}), which trace {\em in situ} stellar mass, and the proposed u-band imaging, which traces the most recent formed stars, in the deep drilling fields. 

\subsubsection{The link between uv and \hi\ morphology} 
\label{s:morph}

At low redshift ($z=0$) there is a clear relation in the outskirts of disk galaxies with low-level massive star-formation linked to the extended \hi\ disk \citep{Thilker05a,Thilker07b,Holwerda12c}. This linked morphology is seen as evidence of the recent build-up in the very outermost outskirts of galaxies. With LSST u-band morphology and \hi\ mass and kinematics (e.g., does this extended UV disk have a lopsided \hi\ profile?) at higher redshift for direct comparison to the local relations, timescales and efficiency of galaxy disk growth can then be explored. 

\subsubsection{Evolution of Galaxy Kinematics;
the Baryonic Tully-Fisher relation from individual and stacked \hi\ Profiles}
\label{s:tf}

The Baryonic Tully-Fisher relation \citep{Tully77} links the optical luminosity of a galaxy to the maximum rotational velocity ($v_{rot}$), measured from \hi\ profiles. With MIGHTEE and LADUMA, these \hi\ profiles will be available for thousands of galaxies. 

LSST optical imaging can contribute the apparent and absolute magnitude of galaxies in these fields, their position angle and especially a measure of their disk inclination, a key correction for the rotational velocity.

In addition to inclinations for TF analysis of individually detected sources we will want PA and inclination of galaxies to optimize stacking \hi\ profiles and attempt to measure line-widths from this stacked \hi\ signal. For interpreting kinematic profiles -- and comparing optical to \hi\ profiles we will want rest-frame optical sizes, morphology and preferably colors of galaxies (i.e. sSFR).
the primary data-set for this would be the deep optical stacks of the LSST Deep Drilling Fields but in order to link the kinematic properties to the star-formation efficiency, u-band data is critical. 
The kinematics of disk galaxies since $z\sim1$ traces the growth, evolution, and settling of disks over the last 7 Gyr.

\subsubsection{Lyman-Break Galaxies (LBGs) / Star-forming Galaxies (SFGs) at $z\sim 1-3$}

The deep u-band observations are crucial for Lyman-Break Galaxies
(LBGs) / Star-forming Galaxies (SFGs) at $z\sim 1-3$. Much of our
understanding of galaxy evolution in the early Universe ($z>4$) comes
from the study of LBGs, systems marked by significant red colors in
bands which straddle the redshifted Lyman limit / Lyman alpha break
\citep[e.g.,][]{Steidel96, Madau95}. Deep imaging surveys have
been extremely successful in identifying, cataloging, and studying
such galaxies \citep[e.g.,][]{Finkelstein15d, Bouwens15}. 
However the limited UV imaging to sample the Lyman-break at
lower redshifts \citep[e.g.,][]{Oesch18, Hathi10} has led
to this ironic twist: We know considerably more about low-luminosity,
distant ($z>4$) LBGs than about brighter, similar LBGs at $z\sim2$. This is
particularly troubling given the peak in the cosmic SFR density at $z\sim2$
\citep[e.g.,][]{Madau14}. The high-luminosity galaxies ($L>L^*$)
are critical in explaining how various feedback mechanisms cause the
halo mass function to differ from the stellar mass or luminosity
function \citep[e.g.,][]{Bower12}.

The best way to identify LBGs/SFGs at $z\sim2$ is through photometric
redshifts based on their deep multi-wavelength imaging data.  Since
the early days of photometric redshifts, inclusion of UV data has
shown to greatly reduce the catastrophic outlier rates, and improve
redshift estimates for galaxies at $z\sim 1-3$ \citep[e.g.,][]{Stanway03,Rafelski15}.  
Additionally, the rest-UV observations are a
vital tracer of dust and star-formation activity in a galaxy by
directly sampling light from hot stars. There is also a huge potential
to perform detailed investigation of LBGs (and Lyman alpha emitting
galaxies) at this epoch through deep ground and space-based
spectroscopy which can probe rest-UV as well as rest-optical
wavelengths. Through extensive study of these galaxies at $z\sim2$, we can
bridge the gap between $z\sim0$ and the very high-redshift samples to study
the epoch of peak SFR density.

With the relative depths of $u$ to $gri$ in the current LSST plans it would mean that $z\sim4$ is still easier to do than $z\sim2-3$.  This is because U depth is key to U-drop and BM/BX object selection typically used to identify galaxies at these redshifts (even with SED determined redshifts, the critical features are essentially the same as those used in Steidel’s original technique). 

As g is about as deep as r in the current LSST  typical cadence, a decent g-dropout ($z\sim4$) selection, even with just the 1st year of the main survey data, so LSST with its current relative exposure depths will not improve on the state of the art, or if it does, will not make full use of the bands longward of U because they are deeper than necessary given the U depth (see also Ferguson+ 2011 LSST white paper). 

One key point about this redshift range is that it is not only cosmic noon, but it is also where the bulk of massive large scale structure starts to assemble. Given the large area (volume) that LSST can survey, selecting these galaxies and studying their clustering will help us identify a properly representative range of structures (proto-groups/clusters) as they start to assemble and go through the star forming to passive red sequence transition that sees most massive cluster cores dominated by passives at $z<1.5$. Enabling deeper U in the five deep drilling fields would give access to $1 Gpc^3$ between $z=1$ and $3$, so a proper census of the Universe at that redshift, not hindered by cosmic variance. The structures we identify in those fields will properly represent how structures are growing over that redshift range.




\section{Concluding Remarks}
\label{s:conclusions}

We propose that the LSST deep drilling fields (XMM-LSS, ECDFS, ELAIS-S1 and COSMOS) to be observed to a comparable depth as the optical, stacked, high-cadence observations with LSST u-band image stacks optimizing for depth, low surface brightness, or angular resolution. 

These deep LSST u-band observations can be instead of or supplemental to u-band coverage of the medium-deep-fast survey by LSST in the optical bands but deep coverage of the deep drilling fields with u-band allows for many different science topics in coordination with SKA pathfinder observations, especially those with the MeerKAT radio telescope. 



\appendix

\include{metrics}






\section{Cadence}
\label{s:cadence}

The exact cadence is not a priority for the proposed science but we imagine a specific cadence strategy would result in optimal transient science using u-band photometry (especially coordinating with the ThunderKAT survey \citep{Fender17}. ThunderKAT aims to sample these fields on timescales from  1 sec upwards, at approximately logarithmic time intervals: e.g. a 12-hour MeerKAT observation will be sampled at 1, 10, 100 and 1000 second intervals. A similar cadence for LSST u-band deep drilling fields as ThunderKAT (modulo night-time observations and time difference) and the optical band cadence would make considerable sense.
\cite{Brandt18} proposes a cadence of every other night for a duration of a decade for all filters to monitor AGN flickering. 

\section{Gray vs Dark Nights}
\label{s:darknights}
Given the priority on depth, dark nights are preferred for the proposed deep u-band observations. Given that this may be impractical, we propose a third reduction of the u-band cadence observations: one that collates the dark sky observations to explore low surface brightness features in the deep fields.

\section{Technical Description}
\begin{footnotesize}
{\it Describe your survey strategy modifications or proposed observations. Please comment on each observing constraint
below, including the technical motivation behind any constraints. Where relevant, indicate
if the constraint applies to all requested observations or a specific subset. Please note which 
constraints are not relevant or important for your science goals.}
\end{footnotesize}

\subsection{High-level description}
\begin{footnotesize}
{\it Describe or illustrate your ideal sequence of observations.}
\end{footnotesize}
The science cases here focus on deep u-band observations of the four selected deep drilling fields, XMM-LSS, ECDFS, ELAIS-S1 and COSMOS.
Agnostic in principle on the cadence, these are ideally paired to the optical observations of these fields. The main goal is to produce stacked images of comparable depth and quality to the optical LSST deep stacks. 
\vspace{.3in}

\subsection{Footprint -- pointings, regions and/or constraints}
\begin{footnotesize}{\it Describe the specific pointings or general region (RA/Dec, Galactic longitude/latitude or Ecliptic longitude/latitude) for the observations. Please describe any additional requirements, especially if there are no specific constraints on the pointings (e.g. stellar density, galactic dust extinction).}
\end{footnotesize}
A single LSST pointing covers the single MeerKAT pointing LADUMA survey (Figure \ref{f:fov1}) and the different pointings of the MIGHTEE coverage of the remaining deep drilling fields. We therefore suggest that the same deep drilling field points are adopted for the u-band observations proposed here.

\subsection{Image quality}
\begin{footnotesize}{\it Constraints on the image quality (seeing).}\end{footnotesize}
Image quality is important for some of the proposed science, but MeerKAT continuum and \hi\ science is done at a nominal resolution of 6" and even during the worst seeing nights, compatible or better resolution can be obtained with LSST.

\subsection{Individual image depth and/or sky brightness}
\begin{footnotesize}{\it Constraints on the sky brightness in each image and/or individual image depth for point sources.
Please differentiate between motivation for a desired sky brightness or individual image depth (as 
calculated for point sources). Please provide sky brightness or image depth constraints per filter.}
\end{footnotesize}
The ultimate aim is to obtain similar ultimate depths for this filter as the stacked observations in the optical bands (griz): 27AB magnitude or better. 

For the proposed science, a regular (nightly or the bi-nightly proposed for AGN monitoring) cadence would result in both a deep and a higher resolution stacked image, the latter being particularly useful for morphological studies of the massive star-formation in these galaxies.

Depth is the priority, necessitating a long total integration time because of the lower throughput of the u-band but regular observations of the deep drilling fields would also allow for a high-resolution (but shallower) image to be constructed from the optimal seeing nights. 

\subsection{Co-added image depth and/or total number of visits}
\begin{footnotesize}{\it  Constraints on the total co-added depth and/or total number of visits.
Please differentiate between motivations for a given co-added depth and total number of visits. 
Please provide desired co-added depth and/or total number of visits per filter, if relevant.}
\end{footnotesize}
The total co-added depth needs to be comparable to other filter co-added image depth. Following the reasoning of \cite{Brandt18}, the 10-year stacked depth would be $m_{AB} \sim 28$. This is over the full lengths of the observatory's operations and a mid-point of $\sim27AB$ is eminently feasible with a deep and a high-resolution stack.

\subsection{Number of visits within a night}
\begin{footnotesize}{\it Constraints on the number of exposures (or visits) in a night, especially if considering sequences of visits.  }
\end{footnotesize}
The number of visits a night is not of consequence for the proposed science, just the depth and to a lesser extent, image quality.

\subsection{Distribution of visits over time}
\begin{footnotesize}{\it Constraints on the timing of visits --- within a night, between nights, between seasons or between years (which could be relevant for rolling cadence choices in the WideFastDeep. 
Please describe optimum visit timing as well as acceptable limits on visit timing, and options in
case of missed visits (due to weather, etc.). If this timing should include particular sequences
of filters, please describe.}
\end{footnotesize}
This white paper specifically asks for companion u-band data to the optical griz observations. In order to secure similar image quality, synchronizing the cadence with the optical observations makes the most operational sense but given the fact that the AGN scintillation program specifically asks for y-band \citep{Brandt18}, this would be incompatible to perform concurrently (either filter has to be installed).

\subsection{Filter choice}
\begin{footnotesize}
{\it Please describe any filter constraints not included above.}
\end{footnotesize}
LSST's u-band. 

\subsection{Exposure constraints}
\begin{footnotesize}
{\it Describe any constraints on the minimum or maximum exposure time per visit required (or alternatively, saturation limits). Please comment on any constraints on the number of exposures in a visit.}
\end{footnotesize}
None for individual integrations, the main goal is a deep integrated stack. 

\subsection{Other constraints}
\begin{footnotesize}
{\it Any other constraints.}
\end{footnotesize}
To minimize overhead and compatibility with the AGN program on these fields, campaigns of u-band observations with a few months concurrent observations with the optical cadence inter-spaced with a y-band campaign on these fields, makes the most sense.

\subsection{Estimated time requirement}
\begin{footnotesize}
{\it Approximate total time requested for these observations, using the guidelines available at \url{https://github.com/lsst-pst/survey_strategy_wp}.}
\end{footnotesize}
Following the logic and estimates of \cite{Brandt18}, the 7.5 month observing season would include 60 u-band visits each month. However, this concurrence requirement is not needed for the photometry we argue in favor of here, just the total $\sim3600$ u-band visits total to obtain better than 28AB photometric limit over 10 years.  In effect, these observations will have to be performed during dark times each month, limiting the useful time available. But for this program is only a 0.1\%  commitment of the full 2.8 million LSST visit time commitment.

\vspace{.3in}

\begin{table}[ht]
    \centering
    \begin{tabular}{l|l|l|l}
        \toprule
        Properties & Importance \hspace{.3in} \\
        \midrule
        Image quality &  2   \\
        Sky brightness & 1 \\
        Individual image depth & 2  \\
        Co-added image depth & 1  \\
        Number of exposures in a visit   & 3  \\
        Number of visits (in a night)  & 3  \\ 
        Total number of visits & 1  \\
        Time between visits (in a night) & 3 \\
        Time between visits (between nights)  &  3 \\
        Long-term gaps between visits & 3 \\
        Other (please add other constraints as needed) & \\
        \bottomrule
    \end{tabular}
    \caption{{\bf Constraint Rankings:} Summary of the relative importance of various survey strategy constraints. Please rank the importance of each of these considerations, from 1=very important, 2=somewhat important, 3=not important. If a given constraint depends on other parameters in the table, but these other parameters are not important in themselves, please only mark the final constraint as important. For example, individual image depth depends on image quality, sky brightness, and number of exposures in a visit; if your science depends on the individual image depth but not directly on the other parameters, individual image depth would be `1' and the other parameters could be marked as `3', giving us the most flexibility when determining the composition of a visit, for example.}
        \label{tab:obs_constraints}
\end{table}

\subsection{Technical trades}

The technical trade is between the y-band and u-band mounted visits. In principle there is little to no issue packing the u-band visits in the dark nights each month, leaving y-band cadence for the remainder. 

\begin{footnotesize} 
{\em 1. What is the effect of a trade-off between your requested survey footprint (area) and requested co-added depth or number of visits?}
\end{footnotesize}
Trade-off between co-added depth and footprint should follow the MeerKAT depth strategy with the deepest u-band observations on the LADUMA field (CDFS DDF) and shallower co-add depths on the MIGHTEE fields (remaining DDF).

\begin{footnotesize} 
{\em 2. If not requesting a specific timing of visits, what is the effect of a trade-off between the uniformity of observations and the frequency of observations in time? e.g. a `rolling cadence' increases the frequency of visits during a short time period at the cost of fewer visits the rest of the time, making the overall sampling less uniform.}
\end{footnotesize}
The specific timing of visits, apart from ensuring dark nights, is immaterial for the u-band observations for the above science. However, this is a consideration for the AGN variability considerations. Less uniform time sampling is perfectly acceptable and possibly preferable to the above science (but detrimental to any variability or transient observations).

\begin{footnotesize} 
{\em 3. What is the effect of a trade-off on the exposure time and number of visits (e.g. increasing the individual image depth but decreasing the overall number of visits)?}
\end{footnotesize}
This would be acceptable for the proposed science, which is predominantly photometry based. 

\begin{footnotesize} 
{\em 4. What is the effect of a trade-off between uniformity in number of visits and co-added depth? Is there any benefit to real-time exposure time optimization to obtain nearly constant single-visit limiting depth?}
\end{footnotesize}
No there is no benefit for real-time exposure time optimization for this program as it is photometry limited.

\begin{footnotesize} 
{\em 5. Are there any other potential trade-offs to consider when attempting to balance this proposal with others which may have similar but slightly different requests?}
\end{footnotesize}
To minimize the number of filter changer between u and y-band and the lack of a pressing need for uniform time cadence (unlike the AGN proposal), the u-band observations can be packed in a couple of campaigns during dark nights. Not all the fields have to observed in the same night either and a spread, optimized for each season is eminently possible and practical.

\section{Performance Evaluation}
\begin{footnotesize}
{\it Please describe how to evaluate the performance of a given survey in achieving your desired
science goals, ideally as a heuristic tied directly to the observing strategy (e.g. number of visits obtained
within a window of time with a specified set of filters) with a clear link to the resulting effect on science.
More complex metrics which more directly evaluate science output (e.g. number of eclipsing binaries successfully
identified as a result of a given survey) are also encouraged, preferably as a secondary metric.
If possible, provide threshold values for these metrics at which point your proposed science would be unsuccessful 
and where it reaches an ideal goal, or explain why this is not possible to quantify. While not necessary, 
if you have already transformed this into a MAF metric, please add a link to the code (or a PR to 
\href{https://github.com/lsst-nonproject/sims_maf_contrib}{sims\_maf\_contrib}) in addition to the text description. (Limit: 2 pages).}
\end{footnotesize}

\begin{itemize}
    \item We require that the LSST DDFs cover the multi-wavelength data available in each of the four fields (e.g., Figure \ref{f:fov2}), similar to the \cite{Brandt18}, but including the radio-continuum and \hi\ fields, observed by MeerKAT. The relevant existing metric is ``NightPointingMetric". 
    \item To optimize photometric-redshift derivation and source characterization for AGNs and galaxies in the DDFs, we would like to achieve a relatively uniform depth across the LSST {\em ugri} filters. This is economically possible for the {\em u-band} with a modest investment of 0.1\% of the LSST mission time.
    Relevant existing metrics include ``AccumulateCountMetric", ``AccumulateM5Metric", ``Coaddm5Metric", and ``CrowdingMagUncertMetric",
\end{itemize}

\vspace{.6in}

\section{Special Data Processing}
\begin{footnotesize}
{\it Describe any data processing requirements beyond the standard LSST Data Management pipelines and how these will be achieved.}
\end{footnotesize}
There is very little new data-processing in the proposed science: it benefits from a total and optimal seeing stack for each of the DDF. We ask for a third stack, one that is optimized for low sky background (i.e. dark nights only).

\section{References}
\bibliographystyle{apj}
\bibliography{Bibliography}

\end{document}